\documentclass[acmsmall,screen,nonacm]{acmart}
\usepackage{tabularx}

\begin{document}

\title{Production-Grade AI Coding System for Client-Side Development}
\author{Ruihan Wang}
\affiliation{
  \institution{Shanghai Jiao Tong University}
  \city{Shanghai}
  \country{China}
}
\affiliation{
  \institution{Xiaohongshu}
  \country{China}
}

\author{Chencheng Guo}
\affiliation{
  \institution{Xiaohongshu}
  \country{China}
}

\author{Guangjing Wang}
\affiliation{
  \institution{Xiaohongshu}
  \country{China}
}

\begin{abstract}
Deploying large language model–based code generation in real-world client-side development remains challenging due to heterogeneous inputs, strict engineering constraints, and complex interaction logic expressed in product requirement documents (PRDs). Existing design-to-code approaches often focus on visual translation or single-shot generation, and struggle to reliably align generated code with production requirements.

This paper presents a production-grade AI coding system designed for client-side development under realistic industrial constraints. The system adopts a structured, multi-stage pipeline that integrates Figma designs, natural-language PRDs, and domain-specific engineering knowledge into explicit intermediate artifacts, enabling controlled planning and incremental code generation. By grounding PRD understanding in concrete UI components, the system improves alignment between product requirements and implementation.

We evaluate the system on proprietary but realistic datasets derived from production client-side projects. Results show that domain-specific adaptation significantly improves PRD understanding accuracy, while end-to-end evaluations demonstrate high UI fidelity and robust implementation of interaction logic in real-world cases. These findings suggest that structured, artifact-centric pipelines provide a practical foundation for production-grade AI coding systems.
\end{abstract}

\maketitle

\section{Introduction}
Despite recent advances in code generation with large language models (LLMs), integrating such capabilities into real-world software development workflows remains elusive~\cite{abbassi2025unveiling}. While existing tools demonstrate promising results in frontend prototyping or personal scripting tasks, they often fall short in production environments, where generated code must align with product requirements, enterprise standards, and runtime constraints~\cite{michelutti2024systematic}. This gap is especially evident in client-side development, an area that combines architectural complexity, platform fragmentation, and strict design fidelity, yet remains underexplored in the current literature.

This paper presents a production-grade AI coding system designed for real-world client-side engineering scenarios. In contrast to prior studies that emphasize generic usability or simplified sandbox scenarios, our system is designed to operate under realistic industrial constraints. In practice, generated code must conform to internal design specifications, implement requirements expressed in natural-language product requirement documents (PRDs), and integrate with existing enterprise engineering practices. Our system accepts two primary inputs: (1) a Figma design artifact that specifies layout structure, styling, component usage, and (2) a PRD describing behaviors facing users and business logic in natural language. The system outputs client-side code that is directly usable within the target mobile codebase and suitable for deployment in a production environment.

Formulating this problem in a production setting gives rise to several practical challenges. Design artifacts provide accurate UI structure but are devoid of behavioral semantics; layout hierarchies and component names alone cannot infer interaction logic. PRDs, in contrast, contain rich semantic descriptions but are often unstructured, incomplete, and ambiguous. These challenges are further amplified in client-side development, where projects span large and deeply nested codebases with complex lifecycle management, platform-specific constraints, and heterogeneous UI frameworks (e.g., SwiftUI, UIKit, Jetpack Compose). Compilation is slow, debugging often requires physical devices, and runtime observability is limited, making error localization and recovery substantially more costly than in server-side or web-based environments.

To address these challenges, we do not treat AI coding as a single-shot generation problem, but rather as a structured, multi-stage engineering pipeline, inspired by stepwise code generation strategies proposed in recent work~\cite{10.1145/3672456}. Our system transforms heterogeneous inputs into canonical intermediate representations, plans execution via a deterministic protocol, and finally synthesizes code incrementally under explicitly defined constraints. Each stage produces persistent artifacts, enabling human review, auditing, and recovery from failure. Compared with open-ended prompting workflows, this design ensures that both reasoning processes and decision outcomes remain verifiable and reproducible.

Building on this pipeline, we introduce several engineering-centric techniques tailored to client-side development. First, we treat PRD decomposition as a domain-specific entity extraction problem inspired by Named Entity Recognition (NER)~\cite{grishman1996message}, mapping functional requirements to executable logic units anchored to concrete UI components. Second, we employ structured context engineering to enrich LLM inputs with design tokens, spacing rules, component usage policies, and project-specific architectural constraints. Third, we support static UI layout validation by aligning rendered code with the design tree extracted from Figma, enabling automated detection of layout mismatches prior to runtime. Together, these techniques bridge the gap between high-level product specifications and production-grade client-side implementations.

Due to the proprietary nature of the system and codebase, our evaluation is conducted on a controlled, internal test suite composed of real PRDs, Figma designs, and production client-side projects. 
We report comparative results before and after PRD decomposition, present two end-to-end execution cases, and analyze internal metrics on layout fidelity and logic correctness. 
Although the evaluation scale is necessarily constrained, the results consistently indicate that, under realistic production settings, structured and artifact-centric generation pipelines substantially outperform free-form prompting workflows in terms of code quality, robustness, and maintainability.

In summary, this paper makes the following contributions:

\begin{itemize}
    \item \textbf{A structured, production-grade AI coding pipeline for client-side engineering.} We present an artifact-centric AI coding system tailored for client-side development in an industrial environment. The system integrates heterogeneous inputs from Figma designs and natural-language PRDs, enforces alignment with internal engineering guidelines through explicit intermediate artifacts, and generates production-ready mobile code with human-in-the-loop validation.
    \item \textbf{PRD decomposition as UI-anchored logic extraction.} We formulate PRD understanding as a UI logic decomposition task inspired by Named Entity Recognition (NER), and propose a domain-specific UI component taxonomy to anchor interaction logic to UI elements.
    \item \textbf{Empirical evaluation on realistic workloads.}
    We conduct controlled evaluations on proprietary datasets derived from real client-side projects, including quantitative analysis of PRD decomposition accuracy, checklist-based UI fidelity assessment, and PRD logic implementation studies. The results demonstrate that the system can reliably support code generation under practical production constraints.
\end{itemize}

\section{System Design}
\subsection{Core Insight}
The core insight of our system is that reliable AI coding in real scenarios should be treated as a multi-stage engineering pipeline, rather than a single-step generation. In our system, user inputs are progressively transformed into a sequence of explicit intermediate representations that describe what should be implemented and how the work should be executed. Code generation is then reduced to executing these representations under well-defined constraints. This separation allows reasoning, validation, and human intervention to occur before irreversible code changes are made, significantly improving controllability and robustness in practice.

\subsection{System Overview}
To support the pipeline described above, the system is organized as a two-tier agentic architecture consisting of a client-side coding agent and a backend capability server. Operationally, the client and server cooperate along the pipeline. The coding agent is responsible for reasoning, code editing, and mediating human-in-the-loop interactions, while the server manages the main system pipeline, including three sequential stages: (1) context canonicalization, which converts heterogeneous inputs, such as Figma and PRD URLs into canonical representations; (2) task planning, which synthesizes an execution plan through a protocol; and (3) execution orchestration, which consumes the plan in a dependency-aware and resumable manner. Each stage produces persistent artifacts that serve as the inputs to the next stage, forming a clear and auditable pipeline. The following sections will describe each stage in detail.

\subsection{Context Canonicalization}
A prerequisite for reliable design-to-code is to convert heterogeneous inputs into a canonical context representation. In our system, this responsibility is handled by a context normalization layer that ingests (1) Figma URLs, (2) PRD, and (3) enterprise engineering standards. The layer outputs structured artifacts that are directly consumable by the downstream planning workflow.

\subsubsection{Figma Canonicalization}
When given a Figma URL, the system first retrieves the corresponding design data via API, and transforms it into a simplified design intermediate representation (IR). The IR captures the design as a typed hierarchical structure, where each node encodes its semantic role, layout geometry, visual appearance, and textual attributes. In addition to the node hierarchy, the IR maintains a global style and token space, in which colors, spacing, typography, and other reusable properties are abstracted into stable identifiers. This separation mirrors the structure of modern client UI frameworks, where layout structure and style tokens are managed independently. 

To further improve robustness across real-world scenarios, the canonicalization process is designed to be fault-tolerant. Optional metadata (such as design variables or auxiliary annotations) is incorporated when available, but its absence does not block the construction of a usable design IR. 

Optionally, the system applies an augmentation step. A rendered preview image of the target design node is analyzed using YOLO as the object detector to identify UI elements~\cite{Redmon_2016_CVPR,diwan2023object}. The detected regions are used to refine the canonical structure by flattening overly deep or redundant hierarchies, inserting detected elements as explicit nodes when appropriate.

Finally, by traversing the hierarchical structure, the system identifies which UI components are involved and summarizes them into a component set. This set enables subsequent stages to ground planning decisions in project-specific component libraries and usage constraints.

\subsubsection{PRD Decomposition}
While the UI framework defines the structural and visual boundaries of a product, it is insufficient to support the behavior of a real system. Without an explicit logic layer, the system can only render interfaces but cannot perform tasks or guarantee correct interactions. UI design mockups are typically accompanied by corresponding PRDs, which enable developers to integrate the software’s visual appearance with its functional logic.

The primary objective of PRD processing is to translate interaction logic described in natural language into executable and well-scoped logical units. These logic units typically exhibit several characteristics: they are largely independent and can be generated separately; they must be explicitly bound to corresponding UI components to avoid misalignment between logic and UI; and in certain cases, they require awareness of surrounding UI context to ensure globally consistent interaction semantics.

In practice, PRD content is provided either as raw text or fetched from document links through an external service. Unlike UI descriptions derived from Figma, PRDs are highly unstructured and expressed in free-form natural language. To mitigate the negative impact of this ambiguity on code generation quality, we design a dedicated PRD ingestion process that converts unstructured PRDs into structured representations using a large language model trained specifically for this purpose (details described in \autoref{sec:finetune}).

Inspired by the paradigm of Named Entity Recognition (NER)~\cite{grishman1996message}, our approach treats UI components as the primary entities to be identified within PRD text. Each extracted entity corresponds to a concrete UI component, and the surrounding textual descriptions define the scope and semantics of its associated logic. Table~\ref{NER} summarizes the conceptual parallels between classical NER tasks and PRD logic decomposition. Anchoring logic extraction around UI component entities offers several advantages: each logic unit is explicitly grounded in existing UI code, reducing spurious reasoning across similar components; subtasks can be generated and validated independently; and the classification schema remains extensible as new interaction patterns or business requirements emerge.

The structured output of PRD ingestion is persisted as a “requirement understanding” artifact and saved to the local workspace. Once this artifact is reviewed and confirmed by the user, it is treated as the single source of truth for downstream execution, preventing silent reinterpretation or unintended rewrites in later stages.

\begin{table*}[htbp]
\caption{Parallels between NER tasks and PRD decomposition}
\begin{center}
\begin{tabular}{|c|c|c|}
\hline
 & \textbf{NER tasks} & \textbf{PRD decomposition}\\
\hline
\textbf{Goal} & Identify entities from text & Identify UI component entities from PRD \\
\hline
\textbf{Category} & e.g., person/location/organization & e.g., button/input box/list \\
\hline
\textbf{Boundary} & Start and end positions of entities & Scope of control logic \\
\hline
\textbf{Relation} & Semantic relations among entities & Interaction relations among UI components \\
\hline
\textbf{Context} & Dependency on linguistic context & Functional context of UI components \\
\hline
\end{tabular}
\label{NER}
\end{center}
\end{table*}

\subsubsection{Knowledge Injection via Retrieval-Augmented Context}
To generate high quality code used in real projects, we require adherence to enterprise engineering standards. Therefore, the context canonicalization layer performs retrieval-augmented context injection by fetching two types of project knowledge: (1) Component usage guidelines and (2) Baseline engineering rules such as spacing systems and resource usage standards. 

This domain knowledge is maintained in an internal knowledge base constructed from documents routinely used in production workflows, including enterprise specifications such as UI guidelines and coding standards. All documents are centrally managed within a dedicated document space, where they are segmented into chunks and encoded into vector embeddings~\cite{gao2024retrievalaugmentedgenerationlargelanguage}. During context canonicalization, relevant documents or passages are retrieved through a hybrid retrieval strategy that combines vector similarity search with keyword matching. This approach enables robust recall of both semantically related and explicitly referenced engineering constraints.

To ensure system robustness, the retrieval process is controlled by a cache flag and supported by a time-bounded cache with a fixed time-to-live (TTL). Caching is essential because documentation retrieval constitutes an external dependency in the execution pipeline. In practice, a TTL of one hour provides a practical trade-off between system responsiveness and knowledge freshness.

\subsection{Task Planning}
The system formalizes planning as a finite-step protocol rather than a one-shot generation process. Planning is modeled as a stateful interaction in which each step has a well defined objective, explicit inputs and outputs, and a clear completion condition.

Conceptually, the protocol operates as a finite state machine (FSM). Each invocation specifies the current step identifier and returns a structured response describing the required action and whether the planning process has terminated. A key invariant is step monotonicity: the current step must be fully completed and its artifacts persisted before the protocol advances. This prevents partial or out-of-order planning and ensures that downstream steps always consume finalized artifacts.

The protocol supports two modes with different termination behavior. In PRD-only mode, the user’s intent is limited to PRD clarification or refinement. So execution stops after producing a validated requirement understanding artifact. In full coding mode, the protocol proceeds through technical planning and produces a structured task specification for downstream execution.

In addition, the system introduces several human-in-the-loop checkpoints. These checkpoints ensure that ambiguities are resolved early and prevent incorrect assumptions from propagating into code generation.

\subsection{Execution Orchestration}
The output of the planning protocol is a Task Intermediate Representation (Task IR). Task IR bridges high-level intent and concrete implementation by representing work as a structured set of tasks with explicit dependencies and persistent state.

Task IR is organized as a hierarchical task tree. Internal nodes represent semantic groupings, while leaf nodes are atomic units executed by the coding agent. Each task carries a stable identifier, an objective in natural language, optional execution context, a task type, and an explicit status field.

Task ordering is governed by a local dependency directed acyclic graph (DAG) defined among sibling tasks. The system validates dependencies and performs topological sorting using Kahn’s algorithm before scheduling, ensuring that execution deadlocks are surfaced at planning time rather than during execution~\cite{kahn1962topological}. The orchestrator exposes the next leaf task to the client agent, which performs the corresponding code modification and reports completion status back to the server. Task status updates are applied incrementally and written back to the persisted Task IR. As a result, the orchestrator itself remains largely stateless beyond the stored artifact, enabling execution to resume deterministically after interruptions or restarts. This design ensures that execution failures surface as explicit task states rather than implicit agent errors. Together, Task IR and its scheduler form a deterministic and robust execution backbone for multi-step code generation.

\section{Implementation Details}
\subsection{Implementation Architecture and Engineering Stack}
The system is implemented as a client–server architecture that decouples planning and orchestration from code editing and model interaction. The backend implements a capability server that exposes structured tools for context canonicalization, task planning, and execution orchestration, while the client hosts a coding agent responsible for invoking these capabilities and applying code changes to the target repository.

The coding agent is integrated into an integrated development environment (IDE) extension that supports mainstream development environments, including Visual Studio Code and JetBrains. Through this extension, the agent communicates with the backend capability server via Model Context Protocol (MCP) to invoke structured tools. The underlying code generation model is configurable, allowing the agent to interface with different large language models (e.g., Claude, Gemini~\cite{claude3-2024,gemini1-2023}) without affecting the system architecture.

On the server side, the system is implemented as a long-running TypeScript service on a Node.js runtime. The server exposes its functionality as a set of callable capabilities through MCP, which standardizes how planning, retrieval, and execution tools are registered and invoked by the client agent. The server can be deployed either as a local standard-I/O tool server or as an HTTP service using server-sent events. These capabilities return structured context objects, planning instructions, or task specifications, rather than directly modifying code, ensuring a clean separation of responsibilities between the server and the client agent.

Intermediate artifacts, including normalized context representations and task plans, are serialized using structured JSON and Markdown formats and persisted in the local artifact workspace of a project. This design enables deterministic execution, auditability, and resumable workflows across multiple agent invocations.

\subsection{Execution Workflow}
\begin{figure}[htbp]
  \centering
  \includegraphics[width=\linewidth,height=\textheight,keepaspectratio]{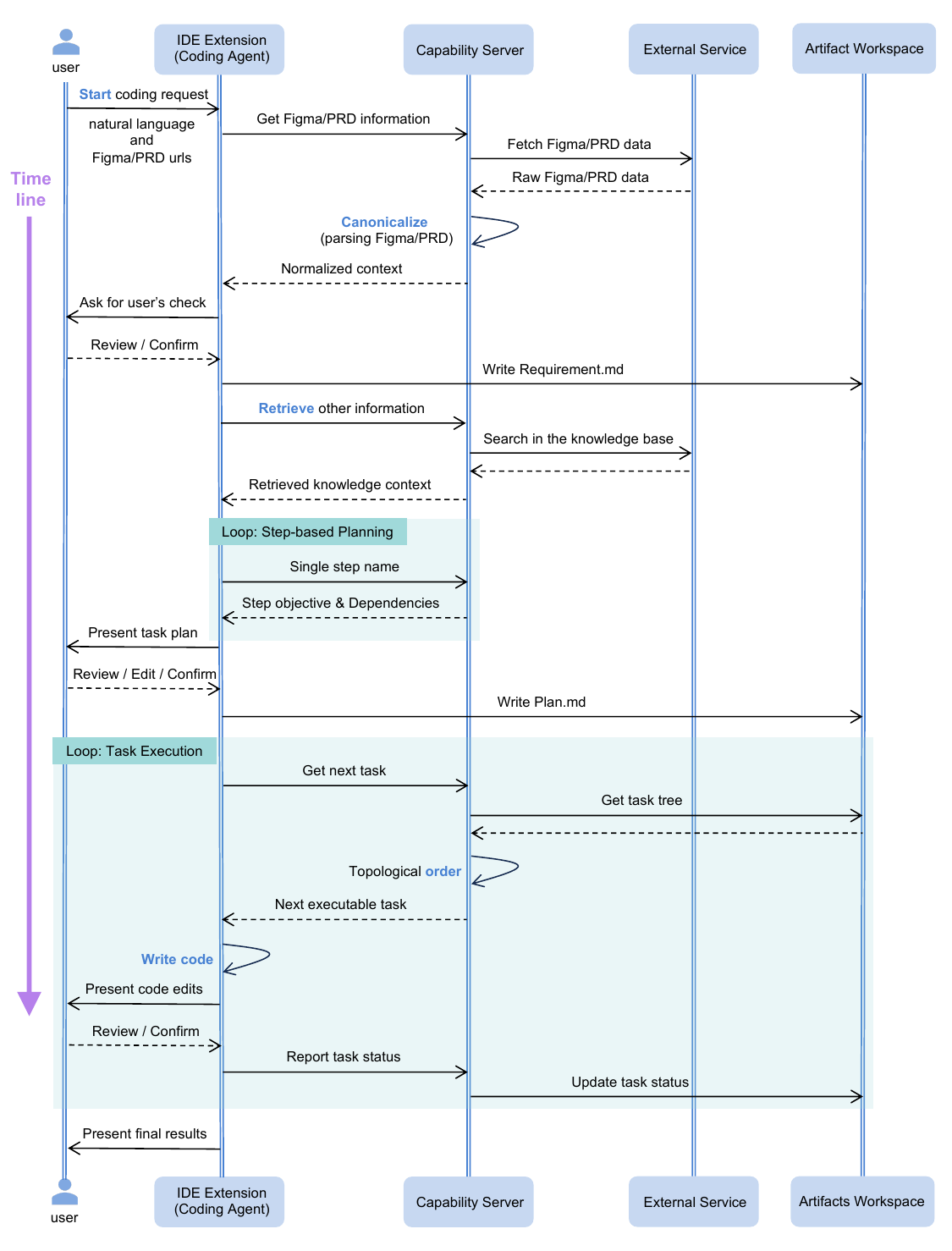}
  \caption{Temporal workflow of the system, showing how natural language and design inputs are progressively transformed into executable code through a sequence of structured stages.
Detailed explanations of the stages and connections are provided in the main text.}
  \label{fig:workflow}
\end{figure}

Figure~\ref{fig:workflow} illustrates the end-to-end execution workflow of the system from the perspective of a user interacting with the coding agent in the full-coding mode. The figure illustrates how user inputs are processed across multiple stages, from context canonicalization and task planning to incremental code execution.

Rather than depicting low-level control flow, the sequence emphasizes the separation between planning and code editing, as well as the role of explicit intermediate artifacts in mediating interactions between the client-side agent and the backend capability server. This design enables controlled and auditable execution while keeping all irreversible code modifications localized to the client environment.

\subsection{Fine-tuning for PRD Decomposition}
\label{sec:finetune}
To enable accurate PRD decomposition, we fine-tune a large language model to identify UI component entities and map them to executable logic units. A critical prerequisite for this process is a well-defined taxonomy of UI components, which serves as the classification space for PRD decomposition and directly determines the granularity and correctness of generated logic tasks.

\subsubsection{Design of UI Component Taxonomy}

The core of PRD decomposition lies in the definition of UI component categories. We establish the taxonomy based on three scientifically grounded criteria.

\textbf{Completeness.} The taxonomy should cover all common UI components encountered in mobile applications. Every component appearing in real-world PRDs must be assignable to at least one category, ensuring both coverage and exhaustiveness.

\textbf{Mutual Exclusivity.} Each UI component should belong to only one primary category. Clear boundaries and explicit classification rules are defined to guarantee unambiguous assignment and avoid overlapping logic responsibilities during code generation.

\textbf{Practical Utility.} Beyond conceptual clarity, the taxonomy must directly support downstream engineering tasks. Each category is designed to correspond to a distinct class of logic generation patterns, thereby providing actionable guidance for task decomposition.

Based on industry design guidelines and practical development experience, we categorize UI components into seven classes, as summarized in Table~\ref{controls}.

\begin{table}[htbp]
\caption{UI control categories}
\begin{center}
\begin{tabularx}{\linewidth}{|c|X|}
\hline
\textbf{UI controls} & \textbf{Example}\\
\hline
Input Controls & input boxes, search bars, keyboard triggers \\
\hline
Functional Button Controls & send button, voice button, plus button, back button, confirm/cancel buttons \\
\hline
Overlay Panel Controls & floating cards, toasts, pop-up windows \\
\hline
Navigation Bar and Page Framework Controls & page background, top navigation bar area \\
\hline
Content Display Controls & avatars, labels, message bubbles, in-page structural components \\
\hline
List Selection Controls & dropdown menus, single-/multi-select lists, tabs \\
\hline
Additional Logic Control Conditions & user state validation, first-time interaction checks, card-triggered states, cross-device trigger conditions \\
\hline
\end{tabularx}
\label{controls}
\end{center}
\end{table}

This taxonomy defines the target output space of PRD decomposition and serves as the supervision schema for fine-tuning the decomposition model.

\subsubsection{PRD Decomposition Dataset}
\label{sec:PRD Decomposition Dataset}
To fine-tune the LLM for PRD decomposition, we manually constructed a domain-specific dataset derived from real-world PRDs. The dataset pairs raw PRD segments with annotated UI component entities and their corresponding logical roles, following the taxonomy defined above.

We constructed two complementary datasets.

\textbf{Text-only dataset.} This dataset follows the Alpaca instruction-tuning format. Each sample consists of three fields:
(1) an instruction that explicitly defines the decomposition task, supported UI component categories (Table~\ref{controls}), output format, and classification rules;
(2) an input containing the raw PRD text; and
(3) an output produced by professional software engineers, in which UI components are identified and mapped to their logic descriptions.

\textbf{Multi-modal dataset.} We additionally constructed a multi-modal dataset in the ShareGPT format. In this dataset, images embedded in PRDs, such as UI mockups or screenshots, are included alongside textual instructions. These visual cues provide supplementary context, enabling more accurate alignment between textual logic descriptions and UI structure.

Each dataset contains 182 samples and is randomly split into training and test sets with an 8:2 ratio.

\subsubsection{Fine-tuning Strategy}
We conducted separate fine-tuning experiments on the training datasets. The Qwen2.5-72B-Instruct model was used as the baseline for text-only settings, while Qwen2.5-VL-72B-Instruct was used for multi-modal settings~\cite{bai2025qwen2}. 

For both experiments, we adopted Low-rank adaptation (LoRA) as the fine-tuning method, which allows efficient adaptation of LLMs without modifying the full set of parameters~\cite{hu2022lora}. During training, we systematically varied key hyperparameters, such as learning rate and LoRA rank. Model performance was evaluated on the held-out test set using the F1 score of PRD decomposition.

The fine-tuning was conducted on an internal training platform, using a unified hardware environment of four H20 GPUs. The best-performing models from both the text-only and multi-modal experiments were selected for deployment. Detailed hyperparameter configurations are reported in Table~\ref{finetuneParam}.

\begin{table}[htbp]
\caption{Parameter settings for the two optimal models}
\begin{center}
\begin{tabular}{|c|c|c|}
\hline
& \textbf{Text-only} & \textbf{Multi-modal}\\
\hline
Baseline & Qwen2.5-72B-Instruct & Qwen2.5-VL-72B-Instruct \\
\hline
Cutoff length & 6000 & 6000 \\
\hline
LoRA rank & 4 & 4 \\
\hline
Learning rate & 1e-4 & 1e-5 \\
\hline
Epochs & 30 & 30 \\
\hline
\end{tabular}
\label{finetuneParam}
\end{center}
\end{table}

\section{Test and Evaluation}
\subsection{Evaluation of PRD Decomposition}
We evaluated the effectiveness of the fine-tuned PRD decomposition models in identifying UI control categories from real-world PRDs. The goal of this evaluation is to assess whether fine-tuning enables the model to reliably align unstructured PRDs with the predefined UI control taxonomy (Table~\ref{controls}), which serves as the foundation for subsequent planning and code generation stages.

\subsubsection{Metrics}
We report precision, recall, and F1 score for UI control category prediction. Precision measures the proportion of predicted categories that are correct, recall measures the proportion of ground-truth categories that are successfully identified, and F1 score provides a balanced summary of both aspects. All metrics are computed at the category level and averaged across the test set.

\subsubsection{Evaluation Setup}
We evaluate four model variants:
(1) a text-only baseline model without domain-specific fine-tuning;
(2) a fine-tuned text-only model;
(3) a multi-modal baseline model; and
(4) a fine-tuned multi-modal model.
All models are evaluated on the same held-out PRD decomposition test set described in \autoref{sec:PRD Decomposition Dataset}.

For each PRD in the test set, models are instructed to decompose the input into UI components and assign each component to one of the predefined UI control categories. We focus exclusively on the correctness of the predicted UI control category labels, rather than downstream logic descriptions, in order to isolate the model’s structural understanding of PRDs.

To further analyze the contribution of visual information, we introduce an additional evaluation setting for the fine-tuned multi-modal model. In this setting, all images embedded in the PRDs (e.g., UI mockups or screenshots) are removed, and the model is evaluated using text-only inputs. This no-image evaluation allows us to isolate the impact of visual cues on PRD decomposition performance, while keeping the model architecture and parameters fixed.

Ground-truth labels are provided by experienced client-side engineers, who manually annotated each PRD with the corresponding UI control categories. Model predictions are matched against these annotations at the category level.

\subsubsection{Results Analysis}
Detailed quantitative results are reported in Table~\ref{prdEval}. It shows that domain-specific fine-tuning leads to substantial performance gains over the baseline models in PRD decomposition. For the text-only setting, fine-tuning improves the F1 score from 0.568 to 0.743. The effect of fine-tuning is even more pronounced in the multi-modal setting. The fine-tuned multi-modal model achieves an F1 score of 0.848, compared to 0.211 for the multi-modal baseline, demonstrating that without task-specific adaptation, visual information alone does not yield reliable PRD decomposition. After fine-tuning, the model is able to effectively integrate visual cues with textual descriptions, resulting in both high precision and recall. Removing visual inputs from the fine-tuned multi-modal model leads to a noticeable performance degradation, with the F1 score dropping from 0.848 to 0.751. While the no-image variant still outperforms the text-only baseline, this result confirms that visual context provides complementary information that contributes meaningfully to accurate UI control category identification.

\begin{table}[htbp]
\caption{PRD decomposition performance on UI control category identification}
\label{prdEval}
\centering
\begin{tabular}{|l|c|c|c|}
\hline
\textbf{Model} & \textbf{Precision} & \textbf{Recall} & \textbf{F1} \\
\hline
Text-only Baseline & 0.506 & 0.685 & 0.568 \\
Text-only Fine-tuned & 0.822 & 0.722 & 0.743 \\
\hline
Multi-modal Baseline & 0.202 & 0.256 & 0.211 \\
Multi-modal Fine-tuned & 0.880 & 0.865 & 0.848 \\
Multi-modal Fine-tuned (No Image) & 0.808 & 0.732 & 0.751 \\
\hline
\end{tabular}
\end{table}

\subsection{UI Fidelity Evaluation}
To evaluate whether the generated UI code faithfully reproduces the intended Figma design, we conduct a UI fidelity study on four real modification requests collected from a production mobile application. Each test case corresponds to a concrete UI change proposal and is backed by an associated Figma design. During evaluation, the input to our coding system is \emph{only} the Figma link. After code generation, the resulting UI is rendered in a real device-like preview environment. Professional client-side engineers then compare the rendered result against the target design and perform structured checks.

\subsubsection{Metrics}
We adopt a checklist-based fidelity score that reflects whether the generated UI meets a set of concrete, visually verifiable constraints. Fidelity is defined as:
\begin{equation*}
\text{Fidelity} = \frac{\#\text{Passed Dimensions}}{\#\text{Total Dimensions}}.
\end{equation*}
Each ``dimension'' is a binary criterion (pass/fail). This formulation intentionally avoids subjective holistic ratings and instead decomposes UI correctness into a set of atomic, auditable criteria, improving reproducibility and interpretability of failure cases.

The checklist covers both page-level structure and element-level rendering quality, organized into four groups:

\begin{itemize}
  \item \textbf{Page framework design.} Whether the overall page structure is implemented using an appropriate layout paradigm (e.g., list-based screens should be realized with list primitives rather than ad-hoc stacking).
  \item \textbf{Element constraint relations.} Spatial constraints among UI elements, including relative positioning (top/bottom/left/right anchoring), alignment consistency, and list-specific spacing such as item margins and vertical gaps.
  \item \textbf{Element geometry.} Size correctness and scaling behavior, including whether sizes are properly adaptive where required, or correctly set to absolute values when the design specifies fixed dimensions. For text elements, we additionally check line-wrapping behavior and ellipsis truncation.
  \item \textbf{Element styling.} Visual style attributes including background color, text color, border color, corner radius, icons, and font styling.
\end{itemize}

To mitigate subjectivity, the fidelity checklist was defined prior to evaluation and applied consistently across all test cases. Each dimension corresponds to a concrete, visually verifiable criterion that can be assessed directly from the rendered UI, rather than from code internals or system metadata. Evaluators performed item-by-item checks based solely on observable discrepancies between the generated UI and the target design.

While expert judgment is still required to determine pass or fail for certain fine-grained styling attributes, the checklist-based formulation constrains such judgment to localized, atomic decisions. This design enables failures to be explicitly attributed to specific UI controls and dimensions, improving auditability and reproducibility compared to holistic or impression-based ratings. We acknowledge that this evaluation does not eliminate subjectivity entirely, but we find that it provides a practical and structured mechanism for assessing UI fidelity in production-oriented scenarios.

\subsubsection{Evaluation Setup}
For each test case, the system generates code from the Figma link, and the output is rendered into a device preview. Engineers evaluate the result by (1) enumerating the relevant UI modules and controls appearing on the screen; (2) applying the checklist on each UI control across applicable dimensions; and (3) recording failures with concrete discrepancy notes. 

\begin{figure}[htbp]
  \centering
  \includegraphics[width=0.7\linewidth]{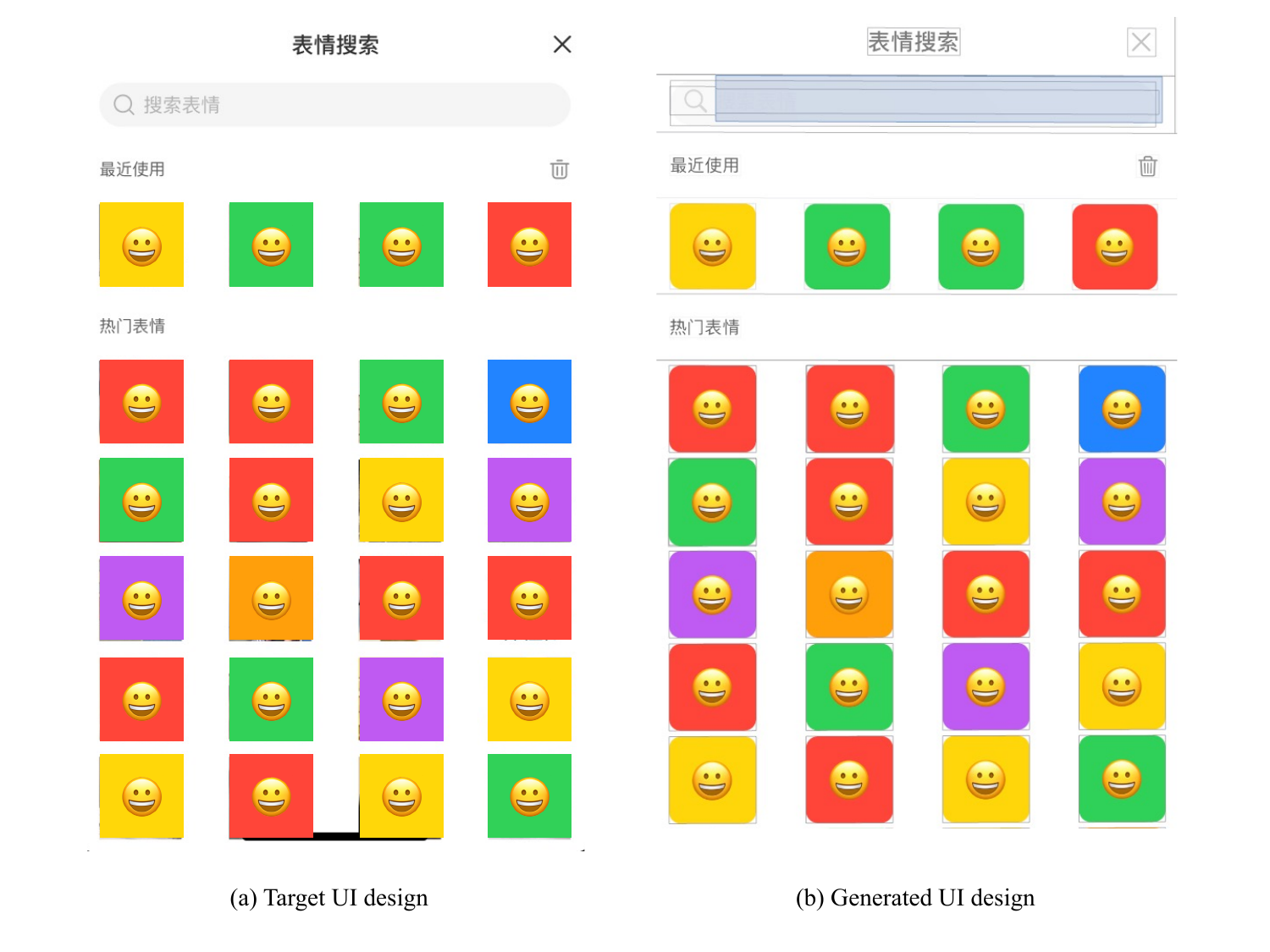}
  \caption{An anonymized example of an emoji searching interface used in our evaluation.
(a) Target UI design. (b) Generated UI design produced by the system.
All visual elements are mock assets and do not correspond to any real product.}
  \label{fig:emoji}
\end{figure}

For instance, in the emoji search case shown in Figure~\ref{fig:emoji}, there are 4 failures out of 36 check items. The failures are localized to specific controls and dimensions, such as incorrect color styling for the close button and placeholder text, a row spacing smaller than designed in the emoji list, and an unintended corner radius on the cell container. Such annotations help distinguish between layout errors and purely stylistic deviations, and provide actionable guidance for iterative correction.

\subsubsection{Results Analysis}
Across four real-world cases, our system achieves consistently high UI fidelity. The results are summarized below:

\begin{itemize}
  \item \textbf{Emoji search:} 89\% fidelity (36 checks, 4 failures).
  \item \textbf{Friend selection:} 89\% fidelity (57 checks, 6 failures).
  \item \textbf{Direct message settings:} 88\% fidelity (34 checks, 4 failures).
  \item \textbf{Profile popup:} 83\% fidelity (18 checks, 3 failures).
\end{itemize}

Overall, the generated UIs match the target designs with fidelity ranging from 83\% to 89\%. A recurring observation is that most failures are \emph{localized} and \emph{non-catastrophic}: they typically involve fine-grained styling attributes, rather than incorrect overall page framework or missing UI modules. This is consistent with the nature of client-side UI implementation, where coarse layout primitives are often recovered correctly from design structure, while pixel-level style and spacing can be sensitive to default component behaviors and platform-specific rendering nuances. These findings suggest that the system is already reliable at reproducing the layout and component structure, and that remaining gaps are concentrated in a small number of easily identifiable visual dimensions, making them amenable to targeted remediation such as stricter style constraints and enhanced design token grounding.

\subsection{PRD Logic Fidelity Evaluation}

In this subsection, we evaluate the system’s ability to correctly implement PRD-specified interaction logic in realistic client-side development scenarios. Unlike UI fidelity, which focuses on visual correctness, this evaluation targets functional requirements that span user interactions and state transitions.

\subsubsection{Metrics}
The evaluated PRD logic includes, but is not limited to:
(1) navigation behaviors (e.g., default tab selection, back navigation);
(2) list interactions (e.g., pull to refresh, pagination, empty state handling);
(3) interaction-triggered state changes (e.g., category filtering and tab selection);
(4) content display rules (e.g., text truncation, timestamp formatting);
and (5) basic robustness behaviors from an engineering perspective (e.g., correct usage of refresh and loading more components).

\subsubsection{Evaluation Setup}
We collect 20 test cases derived from real PRDs collected from a production mobile application. Each case specifies a set of interaction and data logic requirements aggregated over one or more UI components.

During evaluation, the system is provided with both the Figma link and the corresponding PRD document. Code generation is performed under a human-in-the-loop setting, where experienced engineers intervene at several checkpoints to correct obvious errors or ambiguities before final code is produced. This setup reflects the intended usage of the system in practice, where AI-assisted coding operates as a collaborative process rather than a fully autonomous pipeline.

After code generation, professional client-side engineers evaluate generated code at the level of functional correctness (pass/fail), based on whether it fulfills the specified logic and PRD requirements.

\subsubsection{Results Analysis}
Out of the 20 evaluated PRD logic cases, 15 are successfully implemented, resulting in an overall pass rate of 75\%. Importantly, all logic code are generated correctly, indicating that the system is capable of translating PRD descriptions into executable interaction logic under realistic conditions.

The failed cases are primarily attributable to two factors. First, in several instances, the system fails to correctly detect or apply specific framework-level UI components. Second, some failures arise from ambiguous logic descriptions in the generated code, such as missing constraints on text line limits.

Notably, these issues do not reflect a misunderstanding of PRD intent, but rather limitations in component grounding and fine-grained implementation details. Engineers report that such issues are typically localized and can be addressed through improved component retrieval, stronger constraint injection, or tighter integration with UI fidelity detection mechanisms.

\section{Related Work}
\subsection{Design-to-Code Generation}

Automating the transformation from user interface (UI) designs to executable front-end code, commonly referred to as \emph{design-to-code} or \emph{UI-to-code}, has been an active research topic in software engineering for more than a decade. Early efforts primarily relied on computer vision techniques based on rules. Systems such as REMAUI extracted UI components using handcrafted heuristics, optical character recognition, and edge detection, and translated them into platform-specific layout code~\cite{remaui}. While effective for constrained design styles, these methods rely heavily on fixed rules, limiting their robustness to diverse and evolving UI designs.

With the advent of deep learning, subsequent work shifted toward end-to-end neural generation models. Approaches such as pix2code~\cite{beltramelli2018pix2code} and Sketch2Code~\cite{jain2019sketch2code} trained convolutional neural networks to directly map visual UI inputs to executable code~\cite{o2015introduction}. Later models enhanced this paradigm with attention mechanisms or improved visual encoders~\cite{soselia2023learning,chen2022code}, but largely retained a single-shot generation setting. As a result, these methods often struggle with complex layouts, dependencies, and error recovery in realistic scenarios.

To address these limitations, later work introduced multi-stage pipelines that explicitly model UI elements and layout hierarchies. Representative systems decompose the problem into component detection, classification, and layout assembly~\cite{moran2018machine,wu2021screen}, or generate coarse-grained structural representations before filling in detailed code~\cite{gui2025uicopilot,wu2025mllm}. Other approaches segment complex UIs into regions for independent generation~\cite{wan2025divide}, or incorporate iterative refinement through compilation or rendering feedback~\cite{zhou2025declarui}. These methods significantly improve structural correctness and layout fidelity.

In parallel, several datasets and benchmarks have been proposed to support training and evaluation. WebSight provides millions of synthetic screenshot-HTML pairs for scalable model training, while Vision2ui and Web2Code focus on real-world web pages with complex layouts and accompanying metadata~\cite{laurenccon2024unlocking,gui2024vision2ui,yun2024web2code}. Design2Code introduces a curated benchmark with comprehensive automatic and human evaluation metrics, revealing that even state-of-the-art multimodal models frequently produce incomplete or poorly structured code on realistic designs~\cite{si2025design2code}.

Despite these advances, most prior design-to-code research primarily treats code generation as a transformation from visual or partially structured UI representations to implementation. Although recent approaches incorporate hierarchical modeling and region-level decomposition, they generally lack persistent, inspectable intermediate artifacts and do not explicitly integrate higher-level product requirements or interaction logic. These limitations motivate system-level approaches that frame design-to-code as a structured engineering process rather than a purely perceptual or generative task.

\subsection{LLM-based Code Generation}

The emergence of large language models (LLMs) has enabled a new paradigm for code generation. Early demonstrations such as Codex showed that transformer-based models trained on code corpora can generate functional programs from natural language descriptions~\cite{chen2021evaluatinglargelanguagemodels}. AlphaCode extended this direction by combining large-scale sampling with automatic validation to achieve better performance~\cite{li2022competition}. Open-source models, including CodeGen and CodeT5, further explored architectural adaptations for code generation~\cite{nijkamp2022codegen,wang2021codet5identifierawareunifiedpretrained}. More recently, models such as StarCoder and Code Llama have demonstrated that code generation capabilities can scale to large parameter regimes with competitive performance~\cite{li2023starcodersourceyou,roziere2023code}.

Despite these advances at the model level, many LLM-based coding systems still treat code generation as a direct sequence prediction problem. Generated outputs are often syntactically correct but misaligned with user intent or project-specific constraints, particularly in large and evolving codebases. To mitigate these issues, recent work has explored more structured generation paradigms that introduce explicit planning or reasoning steps prior to code synthesis~\cite{10.1145/3672456}. Tree-based reasoning frameworks such as Tree-of-Thought and Tree-of-Code maintain multiple intermediate reasoning or program states to explore alternative solution paths~\cite{yao2023tree,ni2024tree}, while systems such as FunCoder decompose problems into hierarchical plans~\cite{chen2024divide}.

Another emerging direction involves multi-agent systems that simulate collaborative software development process. Frameworks such as MetaGPT and ChatDev organize multiple specialized agents that communicate to design, implement, and validate software artifacts~\cite{hong2023metagpt,qian2024chatdev}. SWE-agent applies similar ideas to real-world GitHub issues by combining retrieval-augmented generation with automated testing~\cite{yang2024sweagent}. Other systems enhance single-agent loops through structured reasoning--action cycles or reflection mechanisms~\cite{shinn2023reflexion,yao2022react}, while Blueprint2Code formalizes multi-agent coordination via persistent planning and execution artifacts~\cite{mao2025blueprint2code}.

Overall, these frameworks mentioned above represent a clear shift toward multi-stage code generation with intermediate reasoning and feedback. However, effectively aligning LLM outputs with product-level specifications, domain-specific constraints, and evolving codebases remains an open challenge. In particular, most existing systems do not ground planning and execution in persistent, inspectable artifacts that explicitly bridge product requirements, design structure, and implementation. These gaps motivate our approach, which integrates explicit intermediate representations and protocolized execution to support production-grade AI coding.

\section{Discussion}

This section discusses the broader implications of our system design choices, analyzes the generalizability and limitations of the proposed approach, and reflects on the role of structured, artifact-centric pipelines in production-grade AI coding systems.

\subsection{Why Multi-stage Pipeline Matters}

A central design choice of our system is to treat AI coding as a multi-stage engineering pipeline rather than a single-shot generation task. In practice, this choice is driven less by model capability limitations than by the operational realities of client-side software development.

In large client-side codebases, errors are expensive to localize and recover from. Under such conditions, normal generation workflows tend to produce brittle outputs: when a generated change fails, developers have limited visibility into the reasoning process that led to the error, making targeted correction difficult. By contrast, our system enforces explicit intermediate representations at each stage, including normalized context artifacts, requirement understanding documents, technical plans, and execution states. These artifacts serve as stable inspection points that allow both humans and automated checks to intervene before irreversible code changes are applied. Our experience suggests that even when the final output is imperfect, the availability of intermediate artifacts significantly reduces rework cost and increases developer trust in AI-assisted workflows.

\subsection{Generalizability and Domain Specificity}
While the system is developed and evaluated within a single organization, many of its core design principles are domain-agnostic. The separation between context canonicalization, planning, and execution; the use of persistent intermediate artifacts; and the formulation of execution as a dependency-aware task graph are applicable to a wide range of AI-assisted software engineering scenarios beyond client-side development.

At the same time, certain components of the system are intentionally domain-specific. The UI component taxonomy, PRD decomposition schema, and engineering rules reflect conventions and practices specific to mobile client development. Rather than viewing this specialization as a limitation, we argue that it is a necessary condition for production readiness. Our results suggest that fully general AI coding systems without incorporating domain knowledge may appear to perform perfectly, but they are difficult to be deployed as trustworthy tools in real‑world scenarios.

This observation points to a broader implication: effective AI coding systems may need to balance general capabilities with domain-specific intermediate representations. Future work could explore how such representations can be modularized or adapted across domains, for example by introducing pluggable taxonomies or configurable planning protocols.

\subsection{Limitations}
Our study has several limitations that warrant discussion. First, due to proprietary constraints, the system source code and the datasets used for PRD decomposition cannot be publicly released. While this restricts direct reproducibility, it reflects a common trade-off in industrial software engineering research. To mitigate this issue, we provide detailed descriptions of the system design, taxonomy, and evaluation protocols to facilitate conceptual replication.

Second, the UI fidelity evaluation relies on expert assessment using a structured checklist rather than fully automated metrics. Although the checklist is designed to minimize subjectivity by decomposing fidelity into atomic, visually verifiable criteria, some degree of human judgment remains unavoidable. 

Finally, the current evaluation focuses on relatively small-scale UI modification tasks. While these cases are representative of real development scenarios, scaling the system to larger feature implementations or long-term maintenance tasks remains an open challenge. Addressing this will likely require tighter integration with testing frameworks, richer dependency analysis, and more advanced change impact modeling.

\subsection{Implications for Future AI Coding Systems}

The findings of this work suggest that progress in AI-assisted software engineering should not be measured solely by end-to-end success rates. Instead, system-level properties such as controllability, auditability, and recovery cost play a crucial role in determining practical usefulness.

By reframing AI coding as a structured engineering pipeline with explicit intermediate artifacts and protocolized execution, our system demonstrates a viable path toward integrating large language models into real-world development workflows. We believe this perspective complements ongoing advances in model capability and offers a foundation for future research on robust, maintainable, and trustworthy AI coding systems.

\section{Conclusion and Future Work}
\subsection{Conclusion}
This paper presents a production-grade AI coding system for client-side development that bridges the gap between research-oriented code generation and real-world engineering practice. By adopting a structured, multi-stage pipeline with explicit intermediate artifacts, the system integrates Figma designs, natural-language PRDs, and domain knowledge to enable controlled planning and incremental code generation. A central contribution is the formulation of PRD decomposition as UI logic extraction, which improves alignment between product requirements and implementation. Evaluations demonstrate improved PRD understanding, high UI fidelity, and robust logic implementation under realistic constraints. These results highlight the effectiveness of artifact-centric system design for practical AI-assisted software engineering.

\subsection{Future Work}
\subsubsection{Codebase knowledge Graph}
A key limitation of conventional retrieval-augmented generation (RAG)~\cite{lewis2020retrieval} lies in an assumption: knowledge is typically represented as unstructured text, which is well suited for design guidelines or engineering documentation, but less effective for reasoning over large, evolving codebases. In practice, effective code modification requires an understanding of historical evolution and prior implementation decisions.

To address this gap, we are exploring the construction of a knowledge graph based on the codebase that models source files, symbols, APIs, dependency relations, and versioned changes as entities. By hybrid search, the system can retrieve both relevant code and semantically related docs. It has the potential to provide more precise and temporally grounded context for AI coding tasks, particularly in scenarios involving legacy code modification.

\subsubsection{Richer Evaluation Protocols}
Finally, future work will focus on expanding the scope and depth of system evaluation. While the current study evaluates PRD decomposition quantitatively and UI fidelity through structured expert checklists, these evaluations are necessarily limited in scale. As the system matures, we aim to conduct longitudinal evaluations across larger codebases and longer development cycles, capturing metrics such as regression frequency, recovery cost, and developer intervention effort.

In addition, we plan to explore partially automated evaluation techniques, such as layout tree comparison, visual similarity metrics, to complement expert assessment. Together, these efforts may enable more scalable and reproducible evaluation of production-grade AI coding systems.

\section{Data Availability}
The system described in this paper is implemented using internal tooling and infrastructure developed within the authors’ organization. Due to proprietary and confidentiality constraints, the source code of the system, including the MCP tools and execution framework, cannot be publicly released.

The datasets used for PRD decomposition and test cases for evaluation are derived from real-world product requirement documents and client-side development tasks from production projects. These materials contain sensitive product information and therefore cannot be shared. Similarly, the fine-tuning process for PRD decomposition was conducted on an internal platform, where training execution is handled by platform-managed infrastructure rather than user-accessible code.

Despite these restrictions, all experimental metrics, evaluation criteria, and decision procedures are explicitly specified to support conceptual replication of the reported results. We believe that this level of methodological disclosure provides meaningful evidence for the claims made in this paper while respecting the constraints of industrial software development environments.

\bibliographystyle{ACM-Reference-Format}
\bibliography{references}

@article{abbassi2025unveiling,
  title={Unveiling inefficiencies in llm-generated code: Toward a comprehensive taxonomy},
  author={Abbassi, Altaf Allah and Da Silva, Leuson and Nikanjam, Amin and Khomh, Foutse},
  journal={arXiv preprint arXiv:2503.06327},
  year={2025}
}

@inproceedings{michelutti2024systematic,
  title={A Systematic Study on the Potentials and Limitations of LLM-assisted Software Development},
  author={Michelutti, Chiara and Eckert, Jens and Monecke, Milko and Klein, Julian and Glesner, Sabine},
  booktitle={2024 2nd International Conference on Foundation and Large Language Models (FLLM)},
  pages={330--338},
  year={2024},
  organization={IEEE}
}

@inproceedings{grishman1996message,
  title={Message understanding conference-6: A brief history},
  author={Grishman, Ralph and Sundheim, Beth M},
  booktitle={COLING 1996 volume 1: The 16th international conference on computational linguistics},
  year={1996}
}

@article{kahn1962topological,
  title={Topological sorting of large networks},
  author={Kahn, Arthur B},
  journal={Communications of the ACM},
  volume={5},
  number={11},
  pages={558--562},
  year={1962},
  publisher={ACM New York, NY, USA}
}

@article{diwan2023object,
  title={Object detection using YOLO: challenges, architectural successors, datasets and applications},
  author={Diwan, Tausif and Anirudh, G and Tembhurne, Jitendra V},
  journal={multimedia Tools and Applications},
  volume={82},
  number={6},
  pages={9243--9275},
  year={2023},
  publisher={Springer}
}

@article{hu2022lora,
  title={Lora: Low-rank adaptation of large language models.},
  author={Hu, Edward J and Shen, Yelong and Wallis, Phillip and Allen-Zhu, Zeyuan and Li, Yuanzhi and Wang, Shean and Wang, Lu and Chen, Weizhu and others},
  journal={ICLR},
  volume={1},
  number={2},
  pages={3},
  year={2022}
}

@article{bai2025qwen2,
  title={Qwen2. 5-vl technical report},
  author={Bai, Shuai and Chen, Keqin and Liu, Xuejing and Wang, Jialin and Ge, Wenbin and Song, Sibo and Dang, Kai and Wang, Peng and Wang, Shijie and Tang, Jun and others},
  journal={arXiv preprint arXiv:2502.13923},
  year={2025}
}

@techreport{gemini1-2023,
  title        = {Gemini 1: A Family of Highly Capable Multimodal Models},
  author       = {Google DeepMind},
  institution  = {Google DeepMind},
  year         = {2023},
  url          = {https://storage.googleapis.com/deepmind-media/gemini/gemini_1_report.pdf},
  note         = {Accessed: 2026-01-28}
}

@misc{claude3-2024,
  author       = {{Anthropic}},
  title        = {The Claude 3 model family},
  howpublished = {\url{https://www.anthropic.com/news/claude-3-family}},
  year         = {2024},
  note         = {Accessed: 2026-01-28}
}

@INPROCEEDINGS{remaui,
  author={Nguyen, Tuan Anh and Csallner, Christoph},
  booktitle={2015 30th IEEE/ACM International Conference on Automated Software Engineering (ASE)}, 
  title={Reverse Engineering Mobile Application User Interfaces with REMAUI (T)}, 
  year={2015},
  volume={},
  number={},
  pages={248-259},
  doi={10.1109/ASE.2015.32}}

@inproceedings{beltramelli2018pix2code,
  title={pix2code: Generating code from a graphical user interface screenshot},
  author={Beltramelli, Tony},
  booktitle={Proceedings of the ACM SIGCHI symposium on engineering interactive computing systems},
  pages={1--6},
  year={2018}
}

@article{jain2019sketch2code,
  title={Sketch2Code: transformation of sketches to UI in real-time using deep neural network},
  author={Jain, Vanita and Agrawal, Piyush and Banga, Subham and Kapoor, Rishabh and Gulyani, Shashwat},
  journal={arXiv preprint arXiv:1910.08930},
  year={2019}
}

@article{o2015introduction,
  title={An introduction to convolutional neural networks},
  author={O'shea, Keiron and Nash, Ryan},
  journal={arXiv preprint arXiv:1511.08458},
  year={2015}
}

@article{soselia2023learning,
  title={Learning UI-to-Code Reverse Generator Using Visual Critic Without Rendering},
  author={Soselia, Davit and Saifullah, Khalid and Zhou, Tianyi},
  journal={arXiv preprint arXiv:2305.14637},
  year={2023}
}

@article{chen2022code,
  title={Code generation from a graphical user interface via attention-based encoder--decoder model},
  author={Chen, Wen-Yin and Podstreleny, Pavol and Cheng, Wen-Huang and Chen, Yung-Yao and Hua, Kai-Lung},
  journal={Multimedia Systems},
  volume={28},
  number={1},
  pages={121--130},
  year={2022},
  publisher={Springer}
}

@article{moran2018machine,
  title={Machine learning-based prototyping of graphical user interfaces for mobile apps},
  author={Moran, Kevin and Bernal-C{\'a}rdenas, Carlos and Curcio, Michael and Bonett, Richard and Poshyvanyk, Denys},
  journal={IEEE transactions on software engineering},
  volume={46},
  number={2},
  pages={196--221},
  year={2018},
  publisher={IEEE}
}

@inproceedings{wu2021screen,
  title={Screen parsing: Towards reverse engineering of ui models from screenshots},
  author={Wu, Jason and Zhang, Xiaoyi and Nichols, Jeff and Bigham, Jeffrey P},
  booktitle={The 34th Annual ACM Symposium on User Interface Software and Technology},
  pages={470--483},
  year={2021}
}

@inproceedings{gui2025uicopilot,
  title={UICoPilot: Automating UI synthesis via hierarchical code generation from webpage designs},
  author={Gui, Yi and Wan, Yao and Li, Zhen and Zhang, Zhongyi and Chen, Dongping and Zhang, Hongyu and Su, Yi and Chen, Bohua and Zhou, Xing and Jiang, Wenbin and others},
  booktitle={Proceedings of the ACM on Web Conference 2025},
  pages={1846--1855},
  year={2025}
}

@article{wu2025mllm,
  title={MLLM-Based UI2Code Automation Guided by UI Layout Information},
  author={Wu, Fan and Gao, Cuiyun and Li, Shuqing and Wen, Xin-Cheng and Liao, Qing},
  journal={Proceedings of the ACM on Software Engineering},
  volume={2},
  number={ISSTA},
  pages={1123--1145},
  year={2025},
  publisher={ACM New York, NY, USA}
}

@article{wan2025divide,
  title={Divide-and-Conquer: Generating UI Code from Screenshots},
  author={Wan, Yuxuan and Wang, Chaozheng and Dong, Yi and Wang, Wenxuan and Li, Shuqing and Huo, Yintong and Lyu, Michael},
  journal={Proceedings of the ACM on Software Engineering},
  volume={2},
  number={FSE},
  pages={2099--2122},
  year={2025},
  publisher={ACM New York, NY, USA}
}

@article{zhou2025declarui,
  title={DeclarUI: Bridging Design and Development with Automated Declarative UI Code Generation},
  author={Zhou, Ting and Zhao, Yanjie and Hou, Xinyi and Sun, Xiaoyu and Chen, Kai and Wang, Haoyu},
  journal={Proceedings of the ACM on Software Engineering},
  volume={2},
  number={FSE},
  pages={219--241},
  year={2025},
  publisher={ACM New York, NY, USA}
}

@article{yun2024web2code,
  title={Web2code: A large-scale webpage-to-code dataset and evaluation framework for multimodal llms},
  author={Yun, Sukmin and Thushara, Rusiru and Bhat, Mohammad and Wang, Yongxin and Deng, Mingkai and Wang, Jinhong and Tao, Tianhua and Li, Junbo and Li, Haonan and Nakov, Preslav and others},
  journal={Advances in neural information processing systems},
  volume={37},
  pages={112134--112157},
  year={2024}
}

@article{gui2024vision2ui,
  title={Vision2ui: A real-world dataset with layout for code generation from ui designs},
  author={Gui, Yi and Li, Zhen and Wan, Yao and Shi, Yemin and Zhang, Hongyu and Su, Yi and Dong, Shaoling and Zhou, Xing and Jiang, Wenbin},
  journal={CoRR},
  year={2024}
}

@article{laurenccon2024unlocking,
  title={Unlocking the conversion of web screenshots into html code with the websight dataset},
  author={Lauren{\c{c}}on, Hugo and Tronchon, L{\'e}o and Sanh, Victor},
  journal={arXiv preprint arXiv:2403.09029},
  year={2024}
}

@inproceedings{si2025design2code,
  title={Design2code: Benchmarking multimodal code generation for automated front-end engineering},
  author={Si, Chenglei and Zhang, Yanzhe and Li, Ryan and Yang, Zhengyuan and Liu, Ruibo and Yang, Diyi},
  booktitle={Proceedings of the 2025 Conference of the Nations of the Americas Chapter of the Association for Computational Linguistics: Human Language Technologies (Volume 1: Long Papers)},
  pages={3956--3974},
  year={2025}
}

@misc{chen2021evaluatinglargelanguagemodels,
      title={Evaluating Large Language Models Trained on Code}, 
      author={Mark Chen and Jerry Tworek and Heewoo Jun and Qiming Yuan and Henrique Ponde de Oliveira Pinto and Jared Kaplan and Harri Edwards and Yuri Burda and Nicholas Joseph and Greg Brockman and Alex Ray and Raul Puri and Gretchen Krueger and Michael Petrov and Heidy Khlaaf and Girish Sastry and Pamela Mishkin and Brooke Chan and Scott Gray and Nick Ryder and Mikhail Pavlov and Alethea Power and Lukasz Kaiser and Mohammad Bavarian and Clemens Winter and Philippe Tillet and Felipe Petroski Such and Dave Cummings and Matthias Plappert and Fotios Chantzis and Elizabeth Barnes and Ariel Herbert-Voss and William Hebgen Guss and Alex Nichol and Alex Paino and Nikolas Tezak and Jie Tang and Igor Babuschkin and Suchir Balaji and Shantanu Jain and William Saunders and Christopher Hesse and Andrew N. Carr and Jan Leike and Josh Achiam and Vedant Misra and Evan Morikawa and Alec Radford and Matthew Knight and Miles Brundage and Mira Murati and Katie Mayer and Peter Welinder and Bob McGrew and Dario Amodei and Sam McCandlish and Ilya Sutskever and Wojciech Zaremba},
      year={2021},
      eprint={2107.03374},
      archivePrefix={arXiv},
      primaryClass={cs.LG},
      url={https://arxiv.org/abs/2107.03374}, 
}

@article{li2022competition,
  title={Competition-level code generation with alphacode},
  author={Li, Yujia and Choi, David and Chung, Junyoung and Kushman, Nate and Schrittwieser, Julian and Leblond, R{\'e}mi and Eccles, Tom and Keeling, James and Gimeno, Felix and Dal Lago, Agustin and others},
  journal={Science},
  volume={378},
  number={6624},
  pages={1092--1097},
  year={2022},
  publisher={American Association for the Advancement of Science}
}

@article{nijkamp2022codegen,
  title={Codegen: An open large language model for code with multi-turn program synthesis},
  author={Nijkamp, Erik and Pang, Bo and Hayashi, Hiroaki and Tu, Lifu and Wang, Huan and Zhou, Yingbo and Savarese, Silvio and Xiong, Caiming},
  journal={arXiv preprint arXiv:2203.13474},
  year={2022}
}

@misc{wang2021codet5identifierawareunifiedpretrained,
      title={CodeT5: Identifier-aware Unified Pre-trained Encoder-Decoder Models for Code Understanding and Generation}, 
      author={Yue Wang and Weishi Wang and Shafiq Joty and Steven C. H. Hoi},
      year={2021},
      eprint={2109.00859},
      archivePrefix={arXiv},
      primaryClass={cs.CL},
      url={https://arxiv.org/abs/2109.00859}, 
}

@misc{li2023starcodersourceyou,
      title={StarCoder: may the source be with you!}, 
      author={Raymond Li and Loubna Ben Allal and Yangtian Zi and Niklas Muennighoff and Denis Kocetkov and Chenghao Mou and Marc Marone and Christopher Akiki and Jia Li and Jenny Chim and Qian Liu and Evgenii Zheltonozhskii and Terry Yue Zhuo and Thomas Wang and Olivier Dehaene and Mishig Davaadorj and Joel Lamy-Poirier and João Monteiro and Oleh Shliazhko and Nicolas Gontier and Nicholas Meade and Armel Zebaze and Ming-Ho Yee and Logesh Kumar Umapathi and Jian Zhu and Benjamin Lipkin and Muhtasham Oblokulov and Zhiruo Wang and Rudra Murthy and Jason Stillerman and Siva Sankalp Patel and Dmitry Abulkhanov and Marco Zocca and Manan Dey and Zhihan Zhang and Nour Fahmy and Urvashi Bhattacharyya and Wenhao Yu and Swayam Singh and Sasha Luccioni and Paulo Villegas and Maxim Kunakov and Fedor Zhdanov and Manuel Romero and Tony Lee and Nadav Timor and Jennifer Ding and Claire Schlesinger and Hailey Schoelkopf and Jan Ebert and Tri Dao and Mayank Mishra and Alex Gu and Jennifer Robinson and Carolyn Jane Anderson and Brendan Dolan-Gavitt and Danish Contractor and Siva Reddy and Daniel Fried and Dzmitry Bahdanau and Yacine Jernite and Carlos Muñoz Ferrandis and Sean Hughes and Thomas Wolf and Arjun Guha and Leandro von Werra and Harm de Vries},
      year={2023},
      eprint={2305.06161},
      archivePrefix={arXiv},
      primaryClass={cs.CL},
      url={https://arxiv.org/abs/2305.06161}, 
}

@article{roziere2023code,
  title={Code llama: Open foundation models for code},
  author={Roziere, Baptiste and Gehring, Jonas and Gloeckle, Fabian and Sootla, Sten and Gat, Itai and Tan, Xiaoqing Ellen and Adi, Yossi and Liu, Jingyu and Sauvestre, Romain and Remez, Tal and others},
  journal={arXiv preprint arXiv:2308.12950},
  year={2023}
}

@article{10.1145/3672456,
author = {Jiang, Xue and Dong, Yihong and Wang, Lecheng and Fang, Zheng and Shang, Qiwei and Li, Ge and Jin, Zhi and Jiao, Wenpin},
title = {Self-Planning Code Generation with Large Language Models},
year = {2024},
issue_date = {September 2024},
publisher = {Association for Computing Machinery},
address = {New York, NY, USA},
volume = {33},
number = {7},
issn = {1049-331X},
url = {https://doi.org/10.1145/3672456},
doi = {10.1145/3672456},
journal = {ACM Trans. Softw. Eng. Methodol.},
month = sep,
articleno = {182},
numpages = {30}
}

@article{yao2023tree,
  title={Tree of thoughts: Deliberate problem solving with large language models},
  author={Yao, Shunyu and Yu, Dian and Zhao, Jeffrey and Shafran, Izhak and Griffiths, Tom and Cao, Yuan and Narasimhan, Karthik},
  journal={Advances in neural information processing systems},
  volume={36},
  pages={11809--11822},
  year={2023}
}

@article{ni2024tree,
  title={Tree-of-Code: A Tree-Structured Exploring Framework for End-to-End Code Generation and Execution in Complex Task Handling},
  author={Ni, Ziyi and Li, Yifan and Yang, Ning and Shen, Dou and Lv, Pin and Dong, Daxiang},
  journal={arXiv preprint arXiv:2412.15305},
  year={2024}
}

@article{chen2024divide,
  title={Divide-and-conquer meets consensus: Unleashing the power of functions in code generation},
  author={Chen, Jingchang and Tang, Hongxuan and Chu, Zheng and Chen, Qianglong and Wang, Zekun and Liu, Ming and Qin, Bing},
  journal={Advances in Neural Information Processing Systems},
  volume={37},
  pages={67061--67105},
  year={2024}
}

@inproceedings{yang2024sweagent,
  title={{SWE}-agent: Agent-Computer Interfaces Enable Automated Software Engineering},
  author={John Yang and Carlos E Jimenez and Alexander Wettig and Kilian Lieret and Shunyu Yao and Karthik R Narasimhan and Ofir Press},
  booktitle={The Thirty-eighth Annual Conference on Neural Information Processing Systems},
  year={2024},
  url={https://arxiv.org/abs/2405.15793}
}

@inproceedings{hong2023metagpt,
  title={MetaGPT: Meta programming for a multi-agent collaborative framework},
  author={Hong, Sirui and Zhuge, Mingchen and Chen, Jonathan and Zheng, Xiawu and Cheng, Yuheng and Wang, Jinlin and Zhang, Ceyao and Wang, Zili and Yau, Steven Ka Shing and Lin, Zijuan and others},
  booktitle={The twelfth international conference on learning representations},
  year={2023}
}

@inproceedings{qian2024chatdev,
  title={Chatdev: Communicative agents for software development},
  author={Qian, Chen and Liu, Wei and Liu, Hongzhang and Chen, Nuo and Dang, Yufan and Li, Jiahao and Yang, Cheng and Chen, Weize and Su, Yusheng and Cong, Xin and others},
  booktitle={Proceedings of the 62nd Annual Meeting of the Association for Computational Linguistics (Volume 1: Long Papers)},
  pages={15174--15186},
  year={2024}
}

@article{shinn2023reflexion,
  title={Reflexion: Language agents with verbal reinforcement learning},
  author={Shinn, Noah and Cassano, Federico and Gopinath, Ashwin and Narasimhan, Karthik and Yao, Shunyu},
  journal={Advances in Neural Information Processing Systems},
  volume={36},
  pages={8634--8652},
  year={2023}
}

@inproceedings{yao2022react,
  title={React: Synergizing reasoning and acting in language models},
  author={Yao, Shunyu and Zhao, Jeffrey and Yu, Dian and Du, Nan and Shafran, Izhak and Narasimhan, Karthik R and Cao, Yuan},
  booktitle={The eleventh international conference on learning representations},
  year={2022}
}

@article{mao2025blueprint2code,
  title={Blueprint2Code: a multi-agent pipeline for reliable code generation via blueprint planning and repair},
  author={Mao, Kehao and Hu, Baokun and Lin, Ruixin and Li, Zewen and Lu, Guanyu and Zhang, Zhengyu},
  journal={Frontiers in Artificial Intelligence},
  volume={8},
  pages={1660912},
  year={2025},
  publisher={Frontiers Media SA}
}

@article{lewis2020retrieval,
  title={Retrieval-augmented generation for knowledge-intensive nlp tasks},
  author={Lewis, Patrick and Perez, Ethan and Piktus, Aleksandra and Petroni, Fabio and Karpukhin, Vladimir and Goyal, Naman and K{\"u}ttler, Heinrich and Lewis, Mike and Yih, Wen-tau and Rockt{\"a}schel, Tim and others},
  journal={Advances in neural information processing systems},
  volume={33},
  pages={9459--9474},
  year={2020}
}

@misc{gao2024retrievalaugmentedgenerationlargelanguage,
      title={Retrieval-Augmented Generation for Large Language Models: A Survey}, 
      author={Yunfan Gao and Yun Xiong and Xinyu Gao and Kangxiang Jia and Jinliu Pan and Yuxi Bi and Yi Dai and Jiawei Sun and Meng Wang and Haofen Wang},
      year={2024},
      eprint={2312.10997},
      archivePrefix={arXiv},
      primaryClass={cs.CL},
      url={https://arxiv.org/abs/2312.10997}, 
}

@InProceedings{Redmon_2016_CVPR,
author = {Redmon, Joseph and Divvala, Santosh and Girshick, Ross and Farhadi, Ali},
title = {You Only Look Once: Unified, Real-Time Object Detection},
booktitle = {Proceedings of the IEEE Conference on Computer Vision and Pattern Recognition (CVPR)},
month = {June},
year = {2016}
}
\end{document}